\documentclass[runningheads]{llncs}
\usepackage[T1]{fontenc}
\usepackage{xcolor}
\usepackage{graphicx}
\usepackage{hyperref}
\usepackage{listings}
\usepackage{color}

\definecolor{lightgray}{rgb}{.9,.9,.9}
\definecolor{darkgray}{rgb}{.4,.4,.4}
\definecolor{purple}{rgb}{0.65, 0.12, 0.82}
\definecolor{white}{rgb}{1.0, 1.0, 1.0}

\lstdefinelanguage{JavaScript}{
  keywords={const, for, let, typeof, new, true, false, catch, function, return, null, catch, switch, var, if, in, while, do, else, case, break},
  keywordstyle=\color{blue}\bfseries,
  ndkeywords={class, export, boolean, throw, implements, import, this},
  ndkeywordstyle=\color{darkgray}\bfseries,
  identifierstyle=\color{black},
  sensitive=false,
  comment=[l]{//},
  morecomment=[s]{/*}{*/},
  commentstyle=\color{purple}\ttfamily,
  stringstyle=\color{red}\ttfamily,
  morestring=[b]',
  morestring=[b]"
}

\begin{document}
\title{Ethics of Software Programming with Generative AI: Is Programming without Generative AI always radical?}
%
%
\author{Marcellin Atemkeng\inst{1} \and
Sisipho Hamlomo\inst{1,2} \and
Brian Welman\inst{1}\and Nicole Oyetunji\inst{1} \and Pouya Ataei \inst{3} \and  Jean Louis K. E Fendji\inst{4}}
\authorrunning{M. Atemkeng et al.}
%
\institute{Department of Mathematics, Rhodes University, Makhanda
\email{m.atemkeng@ru.ac.za}, \email{n.oyetunji@gmail.com}, \email{brianallisterwelman@gmail.com}\\
\and 
Department of Statistics, Rhodes University, Makhanda
\email{s.hamlomo@ru.ac.za}\\
\and
School of Engineering Computer and Mathematical Sciences, Auckland University of Technology, Auckland, New Zealand \\
\email{pouya.ataei@aut.ac.nz}\\
\and Department of Computer Engineering, University Institute of Technology, University of Ngaoundere, Cameroon\\
\email{lfendji@gmail.com} }

\maketitle              
\begin{abstract}
This paper provides a comprehensive analysis of Generative AI (GenAI) potential to revolutionise software coding through increased efficiency and reduced time span for writing code. It acknowledges the transformative power of GenAI in software code generation, while also cautioning against the inherent risks of bias and errors if left unchecked. Emphasising the irreplaceable value of traditional programming, it posits that GenAI is not a replacement but a complementary tool for writing software code. Ethical considerations are paramount with the paper advocating for stringent ethical guidelines to ensure GenAI serves the greater good and does not compromise on accountability in writing software code. It suggests a balanced approach, combining human oversight with AI's capabilities, to mitigate risks and enhance reliability. The paper concludes by proposing guidelines for GenAI utilisation in coding, which will empower developers to navigate its complexities and employ it responsibly. This approach addresses current ethical concerns and sets a foundation for the judicious use of GenAI in the future, ensuring its benefits are harnessed effectively while maintaining moral integrity.

\end{abstract}

\section{Introduction}
Generative AI (GenAI) refers to algorithms and systems capable of generating new content, such as images, text, audio, and videos, that mimic human creativity and intelligence \cite{feuerriegel2024generative,brynjolfsson2023generative,bandi2023power}. These systems use deep learning architectures \cite{lecun2015deep,goodfellow2016deep}, including Generative Adversarial Networks (GANs \cite{creswell2018generative,goodfellow2020generative}), Variational Autoencoders (VAEs \cite{doersch2016tutorial}), and Transformers \cite{lin2022survey,khan2022transformers}, to learn patterns from large datasets and generate novel outputs. ChatGPT, developed by OpenAI, is one of several prominent large language models. It has gained significant public attention for its ability to generate human-like text and engage in various language tasks. Its performance in text generation, comprehension, and adaptability to different contexts has made it a widely discussed model in the field of AI, though it is one among many advanced language models contributing to progress in this area. Its versatility extends to coding, essay writing, and poetry, marking a significant leap in AI technology. Similarly, AlphaCode, a cutting-edge coding assistant, leverages the latest AI advancements to aid developers. It has shown impressive results in code writing, bug detection, and providing optimal programming strategies, earning acclaim from developers for boosting workflow efficiency, automating tasks, and easing the learning curve of new programming languages. AlphaCode is quickly becoming essential for coders aiming to enhance their development process. Google's Bard, powered by LaMDA's transformer technology, debuted in March 2023 to test GenAI's impact on creativity and efficiency. Now rebranded as Gemini, it marks a leap in AI, offering a suite of services from writing aid to intricate problem-solving, showcasing Google's dedication to AI's ethical and user-focused progression. Recent advancements in AI language models, such as OpenAI's GPT-4o and Anthropic's Claude 3.5 Sonnet, have significantly enhanced text generation capabilities across various languages and information streams. These models excel in content creation and offer substantial support to writers, coders, and educators. While GPT-4 has received considerable attention, other models like Claude have also demonstrated impressive capabilities, contributing to the rapid advancement of AI-driven content generation.\\\\
 GenAI has shown considerable promise, yet its implementation is not without ethical concerns and considerations. The integration of such technology necessitates a careful examination of its impact on various aspects of society and individual rights. Several authors have highlighted the ethical considerations of GenAI across various fields of research. The complex challenges of GenAI, such as ChatGPT are highlighted in \cite{dwivedi2023opinion}. Ethical concerns arise due to AI's inability to fully understand legal and ethical norms, leading to potential misuse. Lack of transparency and inherent biases in training data may result in misinformation. GenAI presents complex legal challenges, particularly when it concerns intellectual property rights. There are ongoing uncertainties regarding how copyright, patent, and trademark laws apply to content created by AI. Questions about who holds ownership rights and how these laws intersect with AI-generated works remain largely unanswered \cite{parikh2023empowering}. This means that the provisions for confidentiality ought to be strengthened to prevent the inclusion of sensitive information in the text prompts of AI tools \cite{appel2023generative}. The issue of "hallucinations" in GenAI which can produce incorrect information, is underscored in the work of \cite{brand2023using}. They stress the need for consistent human oversight to ensure the accuracy of content generated by AI. Similarly, \cite{cao2023comprehensive} underlined the importance of GenAI content addressing significant societal challenges to ensure its responsible and beneficial use for the community. This accentuates the important role of reliability and accountability in GenAI systems. \\\\
When examining software programming in light of the current GenAI revolution, it is evident that GenAI offers unprecedented capabilities. GenAI's proficiency in generating code, from snippets to entire modules, has the potential to significantly speed up software programming, reducing time-to-market for software products and enabling developers to focus on more complex, creative tasks \cite{lintula2024exploring,haidabrus2024generative,de2024some}. However, this advancement in technology also introduces ethical concerns that need to be carefully considered to guarantee the ethical use of a portion or full code generated by any GenAI tool \cite{desai2024between,bayer2024legal,ungureanu2023legal}. The question of who holds the rights to AI-generated software code is particularly a concern. We ask how one can appropriately acknowledge GenAI code. Establishing guidelines for the use and recognition of AI-generated software code is crucial to maintaining a fair and equitable creative landscape. Also, there is the potential for bias in AI-generated software code \cite{sherje2024enhancing,ambati2023security}. AI systems are only as unbiased as the data on which they are trained, and if this data contains prejudices, the generated software code may also reflect these biases. GenAI systems also raise privacy concerns because they are trained on vast datasets that may contain sensitive personal data \cite{huang2024genai,eskandani2024towards,kapitsaki2024generative}. When generating software code, these systems typically produce code along with simple examples to illustrate usage. However, as with any AI-generated content, it's important for programmers to review the output carefully. While privacy concerns exist with AI systems in general, major coding assistants are designed with safeguards to avoid including personal or sensitive information in their outputs.\\\\
This paper discusses the benefits of GenAI, like improved efficiency and streamlined coding, which can cut down on development time. However, it also points out the risks, such as the chance of creating biased or incorrect code if algorithms are not properly checked. It considers if programming without GenAI is fundamentally different and suggests that, although traditional methods are valuable, GenAI could lead to new advancements. The paper discusses the ethical issues in using GenAI code, highlighting the importance of transparency and accountability for code generated by AI. It argues that ethical rules are crucial for responsible GenAI code use, which should serve the common good. The paper emphasises the importance of GenAI-generated code and advocates for a balanced approach that combines human oversight with AI, which minimises risks and enhances the reliability of code. The paper proposes a plan with clear rules for using GenAI in software writing, aimed at helping developers grasp its complex parts and use it correctly. It contributes to the debate on ethical AI use in programming and prepares for a future where GenAI tools are used with precision and care.

\section{The Advantages of GenAI Software Codes}
\label{la:AdgenAI}
This section looks at the key benefits of GenAI to programming, focusing on code generation, consistency in code style and structure, enhanced software testing, optimised problem-solving approaches and analysing code.
\begin{itemize}
    \item \textit{Code Generation:}
GenAI supports programmers by generating code fragments. This capability enables developers to save considerable time and effort, as they can focus on higher-level design and problem-solving tasks rather than writing repetitive or boilerplate code \cite{ugare2024improving}. For instance, in the context of dynamic web development, one could utilise GenAI to generate standard code components, such as authentication modules. In addition, GenAI could also be used to create input validation tests against erroneous inputs from users, allowing developers to dedicate their expertise to the most challenging aspects of the project.
\item \textit{Consistency in Code Style and Structure:}
Maintaining a consistent code style and structure across a project is essential for readability, maintainability and collaboration within development teams. GenAI can help enforce coding standards by automatically checking code to ensure compliance with predefined style guidelines \cite{huang2024genai}. For example, it can ensure that indentation is consistent, that variable names are descriptive and follow a consistent pattern, and that functions are appropriately documented. GenAI can also suggest refactoring opportunities to improve the structure of the code. For example, it might recommend splitting a large, complex function into smaller, more manageable modules or replacing repeated snippets of code with reusable functions. 


\item \textit{Enhanced Software Testing:}
GenAI also plays a crucial role in enhancing software testing practices. Traditional testing methods often require substantial manual effort to provide extensive testing coverage. However, GenAI can automatically generate test cases that cover a wide range of testing scenarios, e.g. edge cases that developers might overlook \cite{hnatushenko2024use,de2024some}. In addition, GenAI can analyse code to identify potential bugs or vulnerabilities before compile- or run-time, thereby improving the reliability and security of the software. For example, AI-driven tools can suggest tests that simulate various user behaviours, ensuring the software performs as expected under different conditions. This leads to more robust and reliable software releases.
\item \textit{Optimised Problem-Solving Approaches:}
GenAI can propose optimised solutions for coding challenges. When developers encounter complex problems, GenAI can suggest the best approaches based on its vast knowledge of coding patterns and algorithms \cite{tembhekar2023role,eskandani2024towards,nguyen2023generative}. This feature is particularly valuable in scenarios where multiple solutions are possible, but the optimal one is not immediately apparent. For instance, GenAI can recommend the most efficient data structures and algorithms when optimising a search algorithm, reducing development time and computational resources. This not only helps in overcoming coding obstacles but also contributes to the overall performance and efficiency of the software.
\item \textit{Analysing Code:}
Code analysis is a critical aspect of software development, ensuring that code is not only functional but also efficient, secure, and maintainable. GenAI offers tools to analyse code at various stages of development, providing insights that can significantly enhance the quality and performance of software projects \cite{tembhekar2023role,nguyen2023generative}. By using GenAI, developers can detect inconsistencies, optimise performance, and ensure code quality in a more automated and intelligent manner.
\end{itemize}

\section{The Disadvantages of GenAI Software Codes}
While GenAI software offers powerful capabilities in automating code generation, it also has drawbacks that can affect the results' reliability, transparency and fairness. The following subsections explore these challenges in detail, focusing on the fragility of the generated code, the difficulty of debugging, the need for extensive testing, and the risks associated with inherent biases. Understanding these limitations is critical for developers and organisations who use GenAI code.
\begin{itemize}
    \item \textit{Fragile Outputs:}
GenAI software codes are very sensitive to variations in input prompts. Even minor adjustments to prompt scripts can result in drastically different code output that may not work as intended \cite{hadi2023large,liu2023jailbreaking,sun2023automatic}. This vulnerability requires careful, prompt engineering and iterative refinement. Developers must experiment with different prompt structures and wording to achieve the desired results, which can be time-consuming. This unpredictability poses challenges, particularly in environments where consistent and reliable code is critical, such as production systems or mission-critical applications.
\item \textit{Hard to Debug:}
One of the primary challenges with GenAI code is the difficulty of debugging. GenAI models operate as black boxes, meaning their internal decision-making processes are challenging to interpret. If the generated code does not work as expected, it can be challenging to understand why it does not work and how it can be corrected \cite{liang2024large,vaithilingam2022expectation,nam2024using}. Traditional debugging methods rely on understanding the logic and structure of the code, but with GenAI, this is often opaque. This opacity can lead to longer development cycles as developers spend additional time deciphering the generated code and identifying the root causes of problems. Furthermore, the lack of transparency can also hinder collaboration and knowledge sharing between team members as it becomes difficult to explain the reasons behind the generated code.
\item \textit{Code Need Testing:}
Since the results of GenAI are unpredictable, rigorous testing is essential to ensure reliability and accuracy. Each generated code snippet must be thoroughly tested to verify it works correctly under various conditions. This need for extensive testing increases the development workload and extends the project duration. Automated testing frameworks can help, but they may not capture all edge cases or subtle bugs introduced by GenAI. Ensuring comprehensive test coverage is critical to mitigate the risk of deploying faulty or insecure code.
\item \textit{Bias:}
GenAI relies on large amounts of training data to understand language patterns, and the quality of their outputs is strongly influenced by the data on which they were trained. GenAI can amplify any biases or errors in the training data. For example, the original GPT-4 was initially trained on data up to September 2021, so its suggestions had no knowledge of developments over the years until more recent updates. The generated code may reflect these biases and lead to unfair or discriminatory results \cite{openai_models,grossman2023gptjudge,hacker2024generative}. This problem is of particular concern in applications where fairness and justice are critical, such as hiring algorithms, credit approvals, healthcare, or criminal justice systems. The presence of bias in GenAI results can lead to legal and reputational risks for companies. Biased code can lead to discriminatory practices that may violate legal regulations and damage a company's public image.
\end{itemize}

\section{Is Programming without GenAI always radical?}
The short, overarching and unsatisfactory answer is \emph{yes}. The current momentum and rapid adoption of AI in society and industry utterly wash out any specific demographic's thoughts and concerns about the matter and its impact \cite{Bull2024}. Therefore, from the overall and generalised perspective, opposing this trend inherently makes one radical. The longer and more nuanced answer would be that \emph{it depends}.  As discussed in Section \ref{la:AdgenAI}, there is growing evidence that the use of GenAI within programming through coding assistants has a net positive impact. It is highly beneficial for simple and repetitive tasks, improving the productivity and efficiency of programmers in producing programs. This includes the writing of small, non-critical snippets of code or generating test cases for existing code. Moreover, the pairing of programmers with AI assistants has been shown to significantly speed up technical learning across all levels of computer science expertise \cite{Russo2023Navigating,Peng2023The,Yetistiren2023Evaluating}. 

These statements are substantiated by the community-based StackOverflow Developer Survey (SODS) for 2023 and 2024. In these surveys, respondents were asked to associate benefits with AI tools within the software development process. In 2023, out of 38594 responses, 32.81\% said it increases productivity, 25.17\% said it speeds up learning and 24.96\% claimed it allows them to be more efficient. Following this in 2024 with 36894 responses, these benefits remain the top associated benefits at 81\%, 62.4\% and 58.5\%, respectively, indicating that the majority of AI tool users noticed a significant positive impact on their overall productivity. In the same surveys, respondents were asked what they use AI tools for. In 2023 with 37726 responses, 82.55\% use AI tools for writing code, 48.89\% use it to debug or ask for help, 34.37\% use it to document their code, 30.1\% use it to help them learn a codebase, and 23.87\% use it for testing code. In 2024 with 35978 responses, 82\% of users use it to write code, 67\% use it to search for answers, 56.7\% use it to debug or ask for help, 40.1\% use it for documenting code, 34.8\% use it for generating or synthesising content, 30.9\% use it to help them learn a codebase and 27.2\% use it for testing code. Overall, the benefits have tended towards using GenAI for addressing the tasks as stated previously. Noting that AI sentiment was favourable or very favourable amongst all respondents at 77\% in 2023 and 72\% in 2024, there is a good reason to use GenAI for these often repetitive and monotonous tasks \cite{stackoverflow2023Stack,stackoverflow2024Stack}. 

However, there exist several key concerns. This is best described by examining the responses to the first question provided by the SODS for 2023 and 2024: \textit{``Do you currently use AI tools in your development process?"}. In 2023, out of 89184 responses, 43.78\% said they currently are, 25.46\% said that they are currently not using it, but plan to use it, and 29.4\% said they are currently not using and do not plan to use AI tools. Compared to 2024, where out of 60907 responses, the corresponding percentages to the previous year's answers were 61.8\%, 13.8\% and 24.4\%, respectively. Based on these measures, most participants are already using AI, increasing in 2024 presumably due to those who planned to use it from 2023. However, the number of participants who plan to use AI within programming in 2024 has roughly halved, which indicates a decrease in per-year uptake in the use of this technology. In addition, the percentage difference of respondents who do not currently use and do not plan to use AI tools remains approximately the same with a minor decrease in 2024 \cite{stackoverflow2023Stack,stackoverflow2024Stack}. This is a potential indication of the indifference felt towards the current usefulness and innovations surrounding GenAI within programming.  
\cite{Russo2023Navigating} conducted a thorough qualitative survey amongst 100 screened software engineers from several countries to investigate the adoption, integration and related concerns surrounding GenAI within software development. Amongst several conclusions reached, the major concern with the adoption of GenAI with existing software development workflows came down to compatibility, i.e. how well the AI technology integrates with their programming workflow. In contrast, the study analysed the developer's overall perceptions towards the technology and found that there is no significance between the software engineer's overall perception and the adoption of the AI tools. Despite several participants including concerns about security risks, code quality and ethics, the most prominent opinion towards the adoption of AI in programming came down to whether it fits their current workflow. This indicates the pervasive level of indifference felt at a professional level and the overall neutrality towards its uptake. As shown above, there is indeed a strong indication that GenAI is useful within programming, but several programmers may be indifferent to its usage due to the existing performance and security issues surrounding proprietary AI tools. Namely, GenAI tools are great at simple, repetitive tasks, but do not perform as well when it comes to more complex tasks. 

There exists an insightful demonstration of this provided by the prominent software developer YouTuber, Michael Paulson, known as ``ThePrimeagen" \cite{primeagen2023copilot}. It entails asking the GenAI coding assistants to generate a solution to the famous Quick Sort algorithm using the TypeScript language. In their demonstration, GitHub-Copilot was used and generated the solution provided in listing \ref{lst:gpt_quicksort}. As ThePrimeagen notes in the demonstration, the solution will work and there are no evident errors. However, the original algorithm is designed to be optimised for memory usage and efficient at sorting. In listing \ref{lst:gpt_quicksort}, explicit arrays are created at each recursion at lines 7, 8 and 16. This contradicts the goals of the Quick Sort algorithm and therefore cannot be an accepted implementation. The correct solution is given in listing \ref{lst:human_quicksort}. This view is reflected in the SODS 2024. Of 30661 responses, 66.2\% said the most noticeable challenge with the use of AI in their programming was that they did not trust the output the AI provided. Moreover, 63.3\% said AI tools lacked the context of their codebase, internal architecture and company knowledge, further increasing the probability of these assistants producing irrelevant and unsuitable outputs \cite{stackoverflow2024Stack}. Knowing that the code generated by these AI tools is inherently incorrect requires training and experience in programming and general computer science topics. Therefore, only sufficiently qualified developers would be knowledgeable to pick up on these issues. Furthermore, if there exists a significant risk for these coding assistants to create vulnerabilities or ineffective code in developer workflows, then the potential adoption of such software would only occur in low-risk scenarios, i.e. simple and repetitive tasks \cite{Russo2023Navigating}.

The final concern with adopting the use of GenAI in programming is the impact on programming education and beginner or casual programmers. In computer science and related education programmes, the prevalence of AI tools is high. This has led to several issues with the quality of education these future programmers attain. Since some publicly available large language models have been shown to pass most introductory programming courses, there are concerns about over-reliance on such tools by students. Before GenAI, it was noted that coding knowledge obtained through search engines led to lower rates of recall of the same information. Even though search engines were helpful for professional developers, they negatively impacted the student's critical evaluation and problem-solving skills. With large language models amalgamating this information and tailoring it to the style and requirements of the user, this issue has the potential to have even more detrimental effects on the student and their programming ability. Due to the increased usage of AI tools in programming assignments, educational institutions have shown a motion towards opposing its use by implementing more sophisticated anti-cheating and plagiarism tools or adapting the assignments to make it harder for AI tools to solve them. This can be viewed as contradictory since the goal of programming education is to inform and equip students of methods used in industry \cite{Bull2024}. 

As \cite{Russo2023Navigating} notes, we are currently at the beginning of the era of GenAI tools within programming. Therefore, opinions and insights can be volatile and the opinion of programming without using GenAI is far more nuanced. Sufficient evidence exists to show how useful it can be, but more refinement, improvements and education surrounding GenAI tools are needed. Hence, the answer is that it depends on the ethical guidelines established by educational institutions and industries while programmers are expected to uphold responsible programming practices. Additionally, 

\section{Ethical Concerns in GenAI Code}
One fundamental question that continues to be debated is whether it is ethical to incorporate AI-generated code into software development projects without disclosing its AI-based origins. Consider the following potential reasons that may raise ethical concerns. The software programmer using the GenAI code is presenting the code as if it were their original work. While it may seem problematic when the required acknowledgement is not given when using GenAI code, this practice might not be as unusual as it appears. This type of copying and pasting has been part of software programmers' daily routine even before the advent of generative models. Programmers frequently use code or solutions from various platforms such as Stack Overflow or GitHub's Copilot for AI-assisted coding often without proper acknowledgement or consequences. Sometimes, extensive portions of code are taken from GitHub, or practices such as reusing code from our repositories that were originally written by someone else are employed without proper acknowledgement. These practices are quite common in the programming world and can be seen as part of a collaborative and iterative development process \cite{ahmed2015empirical,kim2004ethnographic}.

Programming is a multifaceted discipline that prospers on collaboration and the cumulative knowledge of the software development community. The belief that every line of code must be originally written is not only unrealistic but also inefficient or ineffective. The true measure of code's worth lies in its ability to perform the tasks it was written for, its robustness against errors, and its readability and reusability. Software programmers are encouraged to effectively leverage existing functions, classes, and libraries, which can save time and resources. This practice is not only logical but also widely encouraged within the programming community. Ethical considerations arise concerning disclosure. Typically, in code reuse, disclosure is inherently part of the code, accessible to those who examine it thoroughly. It is rare for programmers to be obligated to provide a detailed list of all used components unless legal requirements dictate otherwise. However, certain circumstances, such as legal stipulations or specific industry standards, may necessitate such disclosures \cite{gogoll2021ethics}.

The central point of the debate on code reuse is not the act of reusing code, which is a universally acknowledged and fundamental practice in software engineering. Instead, it is about the degree of transparency that should be maintained. Transparency is vital in code reuse, not just for preserving trust among developers but also for proper attribution, particularly when leveraging open-source materials \cite{ahmaro2014taxonomy,capilla2019new}.

The use of GenAI for code generation may raise ethical concerns in specific scenarios, such as certain academic assignments or coding assessments during job interviews. However, its use can be appropriate and even beneficial in other contexts, such as creating Proofs of Concept (POCs) or for learning purposes, as long as it is used transparently and per established guidelines. Clear policies should be specified by educational institutions or employers regarding this matter. If regulations explicitly prohibit the use of GenAI-assisted code without proper attribution, its use would indeed be unethical.  The purpose of these guidelines is often to evaluate a student or programmer's individual problem-solving and coding skills. In educational settings, these guidelines ensure that learning objectives are achieved, and students develop the required competencies \cite{accikgoz2021competency,brandt2020measuring}. During job interviews, the guidelines enable employers to assess a candidate's true abilities to code. Therefore, transparency and compliance with established guidelines are essential to uphold the integrity of both student learning and the assessments. 

It is considered unethical and potentially illegal if the license or terms of use for the GenAI tool mandate disclosure. Compliance with licensing agreements is crucial, as it respects the intellectual property and usage rights set by the creators or providers of software tools. Violating these terms can lead to serious legal consequences and undermine the trust in a company's commitment to ethical practices. 

\section{Responsible Ways to use GenAI Code}
Responsible programming practices are essential in upholding ethical and legal standards and ensuring the integrity of software programming. Responsible programming practices include transparency, accountability, adherence to ethical and legal guidelines, learning from past successes, and acknowledging contributions.
\begin{itemize}
    \item \textit{Transparency:}
The increasing prevalence of GenAI in society has highlighted the importance of incorporating human values in software programming.  Human values, according to \cite{SCHWARTZ19921}, represent individuals' core principles and beliefs, which influence behaviour, decision-making, and interactions. An important aspect of human values in software development is transparency, which involves honesty about decision-making during the development process \cite{obie2023understandingdevelopersperceptionstransparency}. In a preliminary study conducted in  \cite{obie2023understandingdevelopersperceptionstransparency},  developers emphasised the significance of transparency in establishing trust with users and stakeholders, enhancing accountability while cultivating ethical practices. This can be achieved by providing users with clear information about the origin and authenticity of AI-generated content, providing insights on data collection, usage and protection, and implementing industry best practices, adhering to regulatory frameworks designed to ensure responsible AI development. Currently, developers make use of a variety of strategies to improve transparency in software development, this includes conducting thorough investigation and root cause analyses to identify underlying factors which may lead to transparency violations. Subsequently developing corrective action plans, and performing testing and verification to resolve violations \cite{obie2023understandingdevelopersperceptionstransparency}. However, it is important to strike a balance between transparency and ethical considerations. Transparency, if not accompanied by proper safeguards, may inadvertently expose sensitive information or proprietary code, which could result in security vulnerabilities and confidentiality breaches.
\item \textit{Accountability:}
Accountability is essential for the successful operation of organisations, as it involves individuals in every aspect of the organisation taking responsibility for their actions and providing explanations for them \cite{alami2024understandingbuildingblocksaccountability}. In the context of software programming, accountability to its users is particularly crucial. In a study conducted in  \cite{alami2024understandingbuildingblocksaccountability}, two primary forms of accountability for software developers are identified: institutionalised and grassroots. Institutionalised accountability is driven by structural and formal processes such as performance evaluations, while grassroots accountability develops naturally within teams based on peers' and individuals' expectations \cite{alami2024understandingbuildingblocksaccountability}. This encourages a collective sense of responsibility grounded in team norms, and the personal and professional values of programmers \cite{alami2024understandingbuildingblocksaccountability}. 
\item \textit{Ethical Guidelines:}
 Establishing clear regulations, especially in professional and academic institutions, regarding the use of GenAI for code generation is essential. This will enable programmers to maintain the integrity of their work, while complying with intellectual property rights, fostering a culture of ethical practice and promoting fair use. 
Adherence to established ethical guidelines and frameworks, such as the IEEE Global Initiative for Ethical Considerations in AI and Autonomous Systems, can help programmers navigate ethical challenges, and make informed decisions throughout the development lifecycle. By offering resources, recommended practices, and ethical principles, the IEEE Global Initiative equips programmers with the necessary tools to identify, assess and address ethical considerations in their work \cite{ieee_global_initiative}. 
\item \textit{Strategies:}
Responsible practices in programming, particularly when making use of GenAI, can be achieved through a \textit{clear communication}. 
    Establishing clear communication channels and expectations between programmers and stakeholders, to ensure that accountability, transparency and ethics are being upheld throughout the development process. Maintaining \textit{detailed documentation} of the software code which will provide a clear record of decisions made, code changes and test results, which are accounted for. This will, in turn, help identify possible areas of improvement.  Implementing regular \textit{routine evaluations} to assist in determining issues early on, and provide opportunities for feedback and improvement. \textit{Providing educational} resources and training on ethical and legal considerations in software programming, particularly when using GenAI tools, to help programmers understand their responsibilities and the potential impact of their decisions and actions. \textit{Learning from past successes}, looking at case studies and previous instances where ethical challenges were successfully addressed.
\end{itemize}

\section{Conclusion}
In the era of the GenAI revolution, the landscape of software programming is undergoing a significant transformation. GenAI's ability to generate code rapidly accelerates development processes, shortens the time-to-market for software products, and allows developers to allocate more time to intricate and innovative tasks. Despite these advantages, the integration of GenAI raises critical ethical questions, particularly regarding the ownership and proper acknowledgement of AI-generated code. It is imperative to establish clear guidelines for the ethical use and recognition of such code to preserve a just and equitable creative environment. Moreover, the potential for bias in AI-generated code and privacy concerns due to the use of extensive datasets necessitates a cautious approach. Programmers must diligently review AI-generated code to ensure it is free of biases and does not compromise privacy, leveraging the built-in safeguards of advanced coding assistants to protect sensitive information.

This paper provides a comprehensive analysis of GenAI in software code generation. It highlights GenAI's potential to enhance efficiency and reduce software coding time. However, it also addresses the inherent risks, such as the production of biased or flawed code due to inadequate algorithmic oversight. The paper debates the distinctiveness of programming with GenAI compared to traditional methods, acknowledging the value of the latter. Central to the discussion are the ethical considerations surrounding GenAI, emphasising the necessity for transparency in algorithms and accountability for AI-generated code. The paper insists on the importance of ethical guidelines to ensure GenAI's responsible application for the greater good. It advocates for a balanced approach to programming, combining human oversight with AI autonomy to mitigate risks and enhance code reliability. A framework with explicit regulations for employing GenAI in code generation is proposed, aiming to equip programmers with the understanding needed to navigate GenAI's complexities and utilise it effectively. The paper contributes to the ongoing discourse on ethical AI practices in programming, envisioning a future where GenAI is employed with meticulousness and responsibility.

\appendix
\begin{lstlisting}[language=JavaScript, caption={GitHub-Copilot solution to the Quick Sort Algorithm in TypeScript based on the demonstration by \cite{primeagen2023copilot}.}, label={lst:gpt_quicksort}]
// write the quicksort function for me
function qs(arr: number[]): number[] {
  if (arr.length <= 1) {
    return arr;
  }
  const pivot = arr[arr.length - 1];
  const left = [];
  const right = [];
  for (let i = 1; i < arr.length; i++) {
    if (arr[i] < pivot) {
      left.push(arr[i]);
    } else {
      right.push(arr[i]);
    }
  }
  return [...qs(left), pivot, ...qs(right)];
}
\end{lstlisting}

\begin{lstlisting}[language=JavaScript, caption={Correct solution to the Quick Sort Algorithm using TypeScript. Credit: \href{https://github.com/AvraamMavridis/Algorithms-Data-Structures-in-Typescript/blob/master/algorithms/quickSort.md}{github/AvraamMavridis}.}, label={lst:human_quicksort}]
    /**
 * Split array and swap values
 *
 * @param {Array<number>} array
 * @param {number} [left=0]
 * @param {number} [right=array.length - 1]
 * @returns {number}
 */
function partition(array: Array<number>, left: number = 0, right: number = array.length - 1) {
  const pivot = array[Math.floor((right + left) / 2)];
  let i = left;
  let j = right;

  while (i <= j) {
    while (array[i] < pivot) {
      i++;
    }

    while (array[j] > pivot) {
      j--;
    }

    if (i <= j) {
      [array[i], array[j]] = [array[j], array[i]];
      i++;
      j--;
    }
  }

  return i;
}

/**
 * Quicksort implementation
 *
 * @param {Array<number>} array
 * @param {number} [left=0]
 * @param {number} [right=array.length - 1]
 * @returns {Array<number>}
 */
function quickSort(array: Array<number>, left: number = 0, right: number = array.length - 1) {
  let index;

  if (array.length > 1) {
    index = partition(array, left, right);

    if (left < index - 1) {
      quickSort(array, left, index - 1);
    }

    if (index < right) {
      quickSort(array, index, right);
    }
  }

  return array;
}

console.assert(
  quickSort(array, 0, array.length - 1).toString() === '1,1,2,2,6,24,31,32,33,63,123,346,943',
  'Wrong Implementation'
);
\end{lstlisting}

\begin{credits}
\subsubsection{\ackname}

\subsubsection{\discintname}

\end{credits}
%
%
%
 \bibliographystyle{splncs04}
 \bibliography{mybibliography}
%




\end{document}